\documentclass[oneside,english]{amsart}
\usepackage[T1]{fontenc}
\usepackage[latin9]{inputenc}
\usepackage{float}
\usepackage{url}
\usepackage{amstext}
\usepackage{amsthm}
\usepackage{amssymb}
\usepackage{graphicx}
\usepackage[authoryear]{natbib}
\usepackage[foot]{amsaddr}
\usepackage[colorlinks=true,linkcolor=blue,anchorcolor=magenta,citecolor=blue,filecolor=magenta,menucolor=black,runcolor=magenta,urlcolor=blue]{hyperref} \usepackage[ocgcolorlinks]{ocgx2}

\makeatletter

\floatstyle{ruled}
\newfloat{algorithm}{tbp}{loa}
\providecommand{\algorithmname}{Algorithm}
\floatname{algorithm}{\protect\algorithmname}

\numberwithin{equation}{section}
\numberwithin{figure}{section}
\theoremstyle{plain}
\newtheorem{thm}{\protect\theoremname}
\theoremstyle{remark}
\newtheorem{rem}[thm]{\protect\remarkname}

\usepackage{enumitem}

\usepackage{xparse}
\NewDocumentCommand{\ceil}{s O{} m}{%
  \IfBooleanTF{#1} 
    {\left\lceil#3\right\rceil} 
    {#2\lceil#3#2\rceil} 
}

\NewDocumentCommand{\floor}{s O{} m}{%
  \IfBooleanTF{#1} 
    {\left\lfloor#3\right\rfloor} 
    {#2\lfloor#3#2\rfloor} 
}

\usepackage{babel}

\providecommand{\remarkname}{Remark}
\providecommand{\theoremname}{Theorem}

\makeatother

\usepackage{babel}
\providecommand{\remarkname}{Remark}
\providecommand{\theoremname}{Theorem}

\begin{document}
\title{Trimmed Match Design for Randomized Paired Geo Experiments}
\author[Aiyou Chen]{Aiyou Chen, Marco Longfils and Nicolas Remy}
\address{Google LLC}
\email{\{aiyouchen, marcolongfils, nicolasremy\}@google.com}

\date{\today}

\begin{abstract}
How to measure the incremental Return On Ad Spend (iROAS) is a fundamental
problem for the online advertising industry. A standard modern tool
is to run randomized geo experiments, where experimental units are
non-overlapping ad-targetable geographical areas \citep{vaver2011measuring}.
However, how to design a reliable and cost-effective geo experiment
can be complicated, for example: 1) the number of geos is often small,
2) the response metric (e.g. revenue) across geos can be very heavy-tailed
due to geo heterogeneity, and furthermore 3) the response metric can
vary dramatically over time. To address these issues, we propose a
robust nonparametric method for the design, called Trimmed Match Design
(TMD), which extends the idea of Trimmed Match \citep{chen2019robust}
and furthermore integrates the techniques of optimal subset pairing
and sample splitting in a novel and systematic manner.
Some simulation and real case studies are presented. We also point out
a few open problems for future research.
\end{abstract}

\maketitle

\section{Introduction}

How to measure the causal effect of online advertising (e.g. on search,
display, video) is a fundamental problem, not only important to advertisers
but also to ad platforms such as Google and Facebook. Despite lots
of research based on observational studies, see \citet{varian2016causal}
for a recent review, and \citet{sapp2017near}, \citet{chen2018bias}
and references therein for some recent innovations, randomized controlled
experiments, as the gold standard for causal inference \citep{imbens2015causal},
are preferred in order to get unbiased and trustful measurements.
Two kinds of randomized experiments have been proposed: cookie-level experiments,
where a large number of cookies are randomly assigned to two different
ad serving conditions (treatment and control), and geo experiments,
where experimental units are non-overlapping ad-targetable geographical
areas, see \citet{adwords} 
for a list of ``geo targets'' supported by Google.
While a cookie experiment is relatively easier to analyze due to large
sample \citep{johnson2017ghost,gordon2019comparison,kohavi2020trustworthy},
it is mostly used for online metrics (e.g., online conversion) and
its measurement may be inaccurate due to technical
issues such as cross-devices, sign-in/out, cookie churn, etc
(\citet{yen2012host} and \citet{coey2016people}). More importantly, such experiments may be
impossible to execute properly due to data protection laws (e.g. \citet{eu.privacy} in Europe)
and due to increasing restrictions on internet user data collection,
e.g. removal of third-party cookies from web browsers, see
recent updates from \citet{apple.privacy} and \citet{google.privacy}.

To
overcome these issues, geo experiments, first proposed by \citet{vaver2011measuring},
only require collecting the response metric (e.g., revenue) and ad
spend at the geo level, which is an easy task for advertisers and
is free of privacy concerns as there is no need to collect any user data. 
Consequently, randomized geo experiments have become
increasingly more popular as they provide a more rigorous tool for
measuring ad effectiveness.\footnote{See \citet{blake2015consumer}
and examples at \url{https://www.blog.google/products/ads/local-ads/}}

However, geo experiments come with their own challenges. In order to
minimize the spillover effect due to travel across the boundaries
of geos, each geo must be large enough so as to be reasonably isolated
from other geos, which often results in: 1) only a small number
of geos available for experimentation \citep{rolnick2019randomized},
and 2) heavy-tailed
experimental data. Despite randomness in geo assignment, geo
heterogeneity can make it hard to obtain a balanced treatment and
control split especially when the sample size is also small. The situation
can be further complicated by temporal dynamics (e.g., marketing promotions,
a hurricane, covid-19 pandemic)---the split may be balanced earlier but
less so later on, even before the experiment starts. Often there
exists budget constraints (e.g. total ad spend on geos in the treatment
group may be pre-specified, but spends for individual geos are determined
by ad platforms through dynamic auction) which introduces
dependencies on the measurements across geos and violates the independence assumption
made by most statistical models. Lastly, the baseline response for
a geo can be a few orders of magnitude larger than the cookie-level
response while the number of geos is very small relative to the number of cookies,
therefore, in order to detect the same level of incremental
effect, geo experiments can be much more expensive than cookie experiments.

In this paper we consider the design problem regarding randomized
geo experiments, specifically with a matched paired design. The parameter
of interest is the incremental return on ad spend (iROAS), denoted
as $\theta$, which is the ratio of overall incremental sales to overall
incremental ad spend. Let $\mathcal{G}$ be the set of geos in the
population, and for any geo $g\in\mathcal{G}$, let $R_{g}^{T}$ and
$R_{g}^{C}$ be the potential responses for the geo under the two
ad serving conditions, namely treatment ($T$) and control ($C$), respectively,
and similarly define $S_{g}^{T}$ and $S_{g}^{C}$ as potential ad
spends. Then the iROAS $\theta$ can be formally expressed as below:
\begin{align*}
\theta & =\frac{\sum_{g\in\mathcal{G}}(R_{g}^{T}-R_{g}^{C})}{\sum_{g\in\mathcal{G}}(S_{g}^{T}-S_{g}^{C})}
\end{align*}
where the numerator is the overall incremental response and the denominator
is the overall incremental ad spend, but none of them is directly observable.

The primary purpose of the experiment is to estimate $\theta$. With a proper
design, though the actual value of $\theta$ is unknown, we hope to
have a good control of the precision of the estimate before the experiment
is actually run. We follow the hypothesis testing framework:
\begin{align}
H_{0}:\theta=0 & \text{ vs }H_{1}:\theta\geq\theta_{0},\label{eq:hypothesis-testing-framework}
\end{align}
where $\theta_{0}$ is a user-specified \textit{minimum detectable
iROAS}. Given a fixed budget, how shall we come up with a matched paired
design (i.e. a set of paired geos randomly assigned to treatment and
control) so that the experiment data to be collected later would have
the capability to accept $H_{0}$ correctly (when $H_{0}$ holds)
and accept $H_{1}$ correctly (when $H_{1}$ holds), both with high
probabilities, given the challenges mentioned earlier?

Let $N$ be the total number of geos available for experimentation.
Let $R_{g}(t)$ and $S_{g}(t)$ be the observed response and ad spend values respectively,
for geo $g \in \{1, \cdots, N\}$ at time $t$, where $t$ is usually a day or a week.
In some scenarios, the spend value may not exist (e.g. for
new advertisers) before the experiment starts, then $S_{g}(t)$ may
be some proxy (e.g. population size) to inform the relative level
of spend across geos. Both $R_{g}(t)$ and $S_{g}(t)$ are collected
prior to the design, referred to as the pretest data hereafter.

We follow the classical advice (\citet{tukey1993tightening}):
balance what you can, randomize the rest, adjust for the random imbalance
(that is, rerandomization), analyze by planned and balanced double randomization.
As usual, the design procedure here requires the construction of candidate
sets of geo pairs, as well as power evaluation for each candidate
design, see for example \citet{rosenbaum2020design}. However, to the best of
our knowledge, there has not been much research specific to measuring
the ratio of two causal effects such as iROAS yet, see \citet{chen2019robust}
and references therein. Our major contributions include: formulation
of the design problem under a nonparametric iROAS model, and proposal
of the use of Trimmed Match and cross validation (a technique commonly
used in model fitting and machine learning, but not for experimental
design yet) for robust power analysis. Unlike classical optimal design
criteria based on parametric models (\citet{wu2020experiments,box2005statistics})
or Taguchi's robust parameter design \citep{nair1992taguchi}, by
using Trimmed Match and cross validation (CV), our method is nonparametric
and robust against both geo heterogeneity and temporal dynamics. By
cross validation, we split the pretest data into two non-overlapping
time periods, where one is used for constructing matched pairs and
the other for power analysis. If the same data is used for both
pairing and power analysis, it can lead to overestimating the power of
the design due to overfitting.

The rest of the paper is organized as follows: Section 2 provides
some background on geo experiments, and Section 3 describes the iROAS
model and derives a criteria for optimal pairing. The design method
is proposed in Section 4. Section 5 reports some numerical studies
as well as performance comparison with the design based on standard
permutation test. A real case study is reported in Section 6 and
Section 7 concludes with some discussions and open problems.

Trimmed Match Design has been applied for customer studies at Google.
The implementation code in Python is partially available at GitHub \citep{tmcode}.

\section{Background}

To design and analyze randomized geo experiments, \citet{vaver2011measuring}
proposed a two-stage regression model, called geo-based regression
(GBR), where the first stage fits a weighted linear regression to
estimate incremental ad spend, and the second stage fits another weighted
linear regression for the response based on estimated incremental
spend as well as data prior to the experiment. GBR provides a
way to model the iROAS directly and the design can be constructed based
on the variance analysis of the estimator. However, the model is sensitive to the choice
of the regression weights and thus is not robust
against geo heterogeneity, and furthermore, incremental ad spend may
correlate with the residual, despite randomness in the treatment assignment,
which may bias the estimate \citep{chen2019robust}.

To conquer the geo heterogeneity issue, \citep{kerman2017estimating}
proposed a time-based regression (TBR) by aggregating the geo-level
data into the group level (treatment vs control), which may be treated
as a simplified version of the Bayesian structural time series model developed
by \citet{brodersen2015inferring}. TBR belongs to the class of synthetic
control models and requires some untestable assumptions, despite
randomness in geo assignment. Later, based on the TBR model,
\citet{au2018time} proposed a market matching method for designing
geo experiments by maximizing a linear relationship between two time series responses, one for
treatment and the other for control, which requires very few geos (essentially a single
pair of grouped geos) and can be less costly if the TBR model fits
and predicts well. However, the design method only applies to non-randomized
experiments and does not meet the gold standard for causal inference.

Recently \cite{chen2019robust} have proposed a sound statistical framework
for measuring iROAS based on a randomized paired geo experiment and further
developed a robust estimation method, namely Trimmed
Match, which will be described in more details in the next section.

There are also some work regarding how to use both geo-level and cookie-level
data \citep{ye2016seasonality} and geo construction through geographical
clustering \citep{rolnick2019randomized}, which are less relevant
to this paper.

\section{Criteria for Optimal Pairing}

In this section, we first describe the iROAS model for a randomized
paired design and then derive a criteria for optimal pairing.

\subsection{The iROAS model\label{subsec:The-iROAS-model}}

To draw statistical inference of iROAS $\theta$, we adopt the nonparametric
iROAS model proposed by \citet{chen2019robust}, which assumes that
iROAS $\theta$ is the same across geos, an assumption first proposed
by \citet{vaver2011measuring}. Given a randomized design with $n$
pairs of geos, let $Y_{i}$ $(X_{i})$ be the response difference (spend
difference) between the treatment geo and the control geo in the $i$th
pair during the experiment period, for $i=1,\cdots,n$. Then the model
can be stated as follows:
\begin{align}
Y_{i} & =\theta\cdot X_{i}+\epsilon_{i},\text{ where }\epsilon_{i}\text{ is symmetric around 0.}\label{eq:model-iROAS}
\end{align}
Here $X_{i}$ is random. Note that $\epsilon_{i}$ may be correlated
with $X_{i}$, which makes the above model different from traditional
linear models. The main statistical obstacle comes from the fact that
despite symmetry, the distribution of $\{\epsilon_{i}:i=1,\cdots,n\}$
can be very heavy-tailed due to geo heterogeneity.

To illustrate the basic idea behind the model, with some abuse of
notation, let $(R_{ij},S_{ij})$ be the response
and spend during the test period for the $j$th geo in the $i$th pair, where $j\in\{1,2\}$
and $i\in\{1,\cdots,n\}$, and let
\begin{align}
Z_{ij} & =R_{ij}-\theta S_{ij}\label{eq:Z}
\end{align}
which represents the ``uninfluenced'' response in the sense that
the effect due to ad spend has been subtracted. Then if $\theta$
is the true iROAS for each geo, the $Z$ value for a specific geo
would be the same no matter whether that geo would be treated or controlled
(see Lemma 1 of \citet{chen2019robust}).

Let $A_{i}$ indicate the random assignment for the $i$th pair with
$P(A_{i}=1)=P(A_{i}=-1)=1/2$ , where $A_{i}=1$ indicates that the
first geo is assigned to treatment and the second geo to control,
while $A_{i}=-1$ indicates the opposite assignment. By definition
we can rewrite $Y_{i}=A_{i}\cdot(R_{i1}-R_{i2})$ and $X_{i}=A_{i}\cdot(S_{i1}-S_{i2})$,
then the symmetric distribution of $\epsilon_{i}$ in model (\ref{eq:model-iROAS})
follows from the fact that
\begin{align}
Y_{i}-\theta X_{i} & \equiv A_{i}\cdot(Z_{i1}-Z_{i2})\label{eq:diff-Z-symmetry}
\end{align}
where both $Z_{i1}$ and $Z_{i2}$ are invariant to geo assignment.

\subsection{Criteria for optimal pairing\label{subsec:Criteria-for-optimal-design}}

To estimate $\theta$, we adopt the robust method ``Trimmed Match''
\citep{chen2019robust}, which can be described as follows. Since
$\epsilon_{i}(\theta)\equiv Y_{i}-\theta\cdot X_{i}$ follows a symmetric
distribution around 0 at the true $\theta$ value, the Trimmed Match
estimator $\hat{\theta}$ solves the trimmed mean equation below:
\begin{align*}
\frac{1}{n-2\ceil{n\lambda}}\sum_{i=\ceil{n\lambda}+1}^{n-\ceil{n\lambda}}\epsilon_{(i)}(\hat{\theta}) & =0
\end{align*}
where $\epsilon_{(1)}(\theta)\leq\cdots\le\epsilon_{(n)}(\theta)$
are the order statistics of $\{\epsilon_{i}(\theta):i=1,\cdots,n\}$,
and $\lambda\in[0, \overline{\lambda})$ is the trim rate, chosen in a data-driven
manner. Here the constant $\overline{\lambda}$ must be less than 0.5;
in practice we recommend to use 0.25 for post analysis so that no more than half of the data points
are trimmed for the point estimation, while for the design procedure to be introduced later,
we recommend a smaller value (e.g. 0.10) in order to be conservative.
Conceptually, if $\mathcal{I}$ is the set of pairs which
are not trimmed, then
\begin{align*}
\hat{\theta} & =\frac{\sum_{i\in\mathcal{I}}Y_{i}}{\sum_{i\in\mathcal{I}}X_{i}}
\end{align*}
which is easily interpretable: when $\lambda=0$, i.e. no pairs are
trimmed, the point estimate is simply the empirical estimate (i.e.
ratio of the response difference to the spend difference, where the
difference is between treatment and control), otherwise, it is the
empirical estimate based on untrimmed pairs $\mathcal{I}$, which
is sensible as ``outlier'' pairs have been trimmed in an unbiased
manner.

By following the principle of the classical ``alphabetic'' optimal
design criteria \citep{wu2020experiments}, we may use the variance
of $\hat{\theta}$ as the optimization criteria here, i.e. from a
few valid candidate designs, choose the one which minimizes $var(\hat{\theta})$.
Unfortunately, there is no simple closed form for $var(\hat{\theta})$,
since $\lambda$ is unknown and needs to be chosen in a data-driven
manner, and furthermore, when there is a budget constraint, it introduces
some complex dependence in the measurements across geos which makes
the exact form of $var(\hat{\theta})$ intractable. On the other hand,
given a fixed total budget $B$, which is the spend difference between
treatment and control, i.e.
\begin{align*}
\sum_{i=1}^{n}X_{i} & \equiv B,
\end{align*}
by assuming the final pairs are in good quality such that the conservative
choice $\lambda=0$ is a good approximation, then $\hat{\theta}$
simplifies to the empirical estimate and we have
\begin{align*}
var(\hat{\theta}) & = \frac{1}{B^2} \sum_{i=1}^{n}\mathbb{E}(Y_{i}-\theta X_{i})^{2}.
\end{align*}
According to (\ref{eq:diff-Z-symmetry}), the right hand side is equal to
$\frac{1}{B^2} \sum_{i=1}^{n}\big|Z_{i1}-Z_{i2}\big|^{2}$. This
gives a conservative variance estimate, and without proper trimming as
Trimmed Match does, this variance estimate may be sensitive to the existence of outlier
pairs. Nevertheless, this additive form motivates the method
of optimal pairing, to be described in Section \ref{subsec:optimal-pairing}.
To be less sensitive to outlier pairs, partially motivated by the
idea of Lasso regression \citep{tibshirani1996regression}, we propose
the $L_{1}$-loss function defined below as the criteria for optimal
pairing:
\begin{align}
L & =\sum_{i=1}^{n}\big|Z_{i1}-Z_{i2}\big|.\label{eq:optimal-matching-criteria}
\end{align}
That is, for a fixed $n$, it is desirable to choose the set of $n$ matched
geo pairs which minimizes $L$. Of course the $L$ value is unknown
at the design phase, and needs to be evaluated.

\section{The design procedure\label{sec:The-design-procedure}}

Our design procedure consists of two major steps: 1) to construct
an optimal set of $n$ geo pairs based on estimated $L$ for each
possible $n$, and 2) to perform the power analysis for each candidate
design. The final design is chosen from the candidates generated by
step 1 which meets the qualification of power analysis in step 2.
Some marketing restrictions may apply for the final decision. To obtain
robust and cost-effective power analysis, our method is built on top
of Trimmed Match and cross validation.

\subsection{Construction of an optimal subset of $n$ geo pairs ($n\protect\leq \floor{N/2}$)\label{subsec:optimal-pairing}}

Here we look for the pairings that minimize the criteria $L$ defined by (\ref{eq:optimal-matching-criteria}).
Note that $L$ can be considered as the sum of within-pair distances.
When the distance between any two geos is known, this is a combinatorial problem \citep{papadimitriou1998combinatorial}.

Suppose that $D$, an $N\times N$ symmetric matrix, represents the
distances between any two geos. The problem of constructing $\floor{N/2}$
pairs out of the $N$ geos such that the total distance for the
pairs is minimized, is called non-bipartite matching (or maximum matching)
in the literature of operations research. It is easy to verify that
there are $\frac{N!}{(N-2n)!(2n)!!}$ different ways to form $n$ pairs
from a set of $N$ geos, which is impractical to exhaust for large
$N$. Fortunately, when  $n=\floor{N/2}$, fast algorithms have been developed with polynomial
time \citep{edmonds1965maximum,papadimitriou1998combinatorial}, see
\citet{kolmogorov2009blossom} for some recent advancement.

To consider $n<\floor{N/2}$, this requires the removal of $N-2n$
geos. We can expand the distance matrix $D$ from $N\times N$ to
$(2N-2n)\times(2N-2n)$ by introducing $N-2n$ pseudo geos, where
the pairwise distances between pseudo geos are infinite but the distance
from any pseudo geo to any of the $N$ real geos is 0. Then applying
non-bipartite matching to the expanded distance matrix gives us $N-n$
pairs, where $N-2n$ of them contains the pseudo geos, and the remaining
$n$ pairs is the optimal solution. This technique
was first discovered by \citet{lu2001matching}, see Chapter 12 of \citet{rosenbaum2020design}
for a more comprehensive review.

The next problem is how to define $D_{gg'}$, the distance between
geo $g$ and $g'$. The variance discussion in Section \ref{subsec:Criteria-for-optimal-design}
suggests that $D_{gg'}$ should measure the squared
difference of ``uninfluenced responses'' between geo $g$ and $g'$.
Recall that by (\ref{eq:Z}), $Z_{ij}=R_{ij}-\theta S_{ij}$. Also
note that the magnitude of $\theta|S_{i1}-S_{i2}|$ is usually expected
to be much smaller than $|R_{i1}-R_{i2}|$ for practical geo experiments.
Therefore, we may approximate $|Z_{i1}-Z_{i2}|^{2}$ by $|R_{i1}-R_{i2}|^{2}$
based on pretest data. Let $\mathcal{T}_p$ be the time period of data to be used
for geo pairing. Suppose that a subset time period $T\subset\mathcal{T}_{p}$
may represent the test period, say with similar seasonality and similar
level of responses but with possibly different duration, then the squared difference can be approximated
by $\big|\sum_{t\in T}R_{g}(t)-\sum_{t\in T}R_{g'}(t)\big|^{2}$ up to a scale factor.
Of course, we may take multiple time periods in order to get a more
reliable estimate. According to the $L$ criteria (\ref{eq:optimal-matching-criteria}),
this leads to the distance between any two geos $g$ and $g'$ as follows:
\begin{align}
D_{gg'} & \propto\sqrt{\sum_{T\subset\mathcal{T}_{p}}\big|\sum_{t\in T}R_{g}(t)-\sum_{t\in T}R_{g'}(t)\big|^{2}}.\label{eq:geo-distance}
\end{align}
Given a candidate pairing of the $2n$ geos, using the same notation as in (\ref{eq:optimal-matching-criteria}),
the $L$ loss can then be expressed as
\begin{align*}
L & \propto\sum_{i=1}^{n}\sqrt{\sum_{T\subset\mathcal{T}_{p}}\big|\sum_{t\in T}R_{i1}(t)-\sum_{t\in T}R_{i2}(t)\big|^{2}}.
\end{align*}

In practice, it may be hard to find a time period $T$ to be representative
of the test period, nevertheless the $L$ criteria discussed above
provides some guidance. As a rule of thumb, if the test period is
a few weeks, one may take $T$ as a week instead of a day, as the
weekly data absorbs the day-of-the-week effect and thus can be less
noisy; Similarly, if the test period is a few months, one may take
$T$ as a month, to absorb the noise in a more granular scale than a month.
One may optimize the choice based on a power analysis, to be described
in Section \ref{subsec:Power-analysis}. See the numerical studies
in Section \ref{sec:simulations} for more discussions.
\begin{rem}\label{rem:rank-based-pairing}
The $L$ criteria described above may be sub-optimal if the pairing
quality measured by $L$ using data in $\mathcal{T}_{p}$ differs
from the test period dramatically. Extra geo features may be incorporated
into the $L$ criteria to improve the reliability of geo matching.
Besides the $L$ criteria, one may also consider some alternative method,
for example, ranking the geos (e.g. using geo sizes), and then pairing
geos based on their ranks and selecting the pairs based on some measure of the
difference (e.g. revenue difference)
between the two geos within each pair. The ranking method is simple but can be
quite effective especially when the geos are well separated from each
other w.r.t. the response values.
See \citet{stuart2010matching,rosenbaum2020design} for the literature
review regarding various general-purpose matching methods.
\end{rem}

\subsection{\label{subsec:Power-analysis}Power analysis with cross validation}

Given a candidate set of $n$ pairs which has been obtained as above
for a pre-specified $n$, this subsection describes how to perform
a robust power analysis. The key idea is to use an evaluation period
$\mathcal{T}_{e}$ which does not overlap with the pairing period
$\mathcal{T}_{p}$. Suppose that $\mathcal{T}_{e}$ is similar to
the test period in terms of the distribution of the pair-level "uninfluenced response" difference.
For $i\in \{1, \cdots, n\}$ and $j\in \{1, 2\}$,
let $R'_{ij}$ be the baseline response value during
$\mathcal{T}_{e}$ for the $j$th geo in the $i$th pair and let $S'_{ij}$
be the spend proxy for the $j$th geo in the $i$th pair, to be used
for simulating the spend data.

To consolidate the idea, let's consider a so-called hold-back experiment where
the spend in the control geos will be 0, i.e.
the control group gets no spend. We simulate the potential ``treated''
spends by assuming they are proportional to the corresponding spend
proxy. Consider the two geos in the $i$th pair and without loss of
generality, assume that $A_i = -1$, i.e. the first geo is assigned to control, and
the second geo to treatment. Assuming that the iROAS $\theta$ is the same across geos as
in Section \ref{subsec:The-iROAS-model}, then the spend and response data can be generated by:
$S_{i1}=0$ and $R_{i1}=R'_{i1}$ for the control geo, and
$S_{i2}=r\cdot S'_{i2}$ and  $R_{i2}=R'_{i2}+\theta\cdot r\cdot S'_{i2}$ for the treatment geo.
The spend difference and response difference for the $i$th pair can be
calculated as $S_{i2} - S_{i1}$ and $R_{i2} - R_{i1}$, respectively.
Here the parameter $r$ is chosen to meet the budget condition.

The estimation error measured by root mean square error (RMSE) for
a candidate design is evaluated with simulated experiment data, which
is described in details in Algorithm \ref{alg:RMSE-evaluation}.\footnote{It is
desirable to exclude some of the assignments due to random imbalance,
to be consistent with the assignment generation procedure described
later, otherwise the estimate of RMSE may be conservative.} The RMSE evaluation
for a different scenario (e.g. a heavy-up
experiment, \cite{bates2011paid}) can be simulated similarly, but the detail is omitted for
conciseness.

\begin{algorithm}
\caption{\label{alg:RMSE-evaluation}RMSE evaluation for a hold-back experiment
with a fixed budget}

\begin{itemize}
\item Input: budget $B$, iROAS $\theta$ (0 under $H_{0}$ and $\theta_{0}$
under $H_{1}$), and data during $\mathcal{T}_e$ for $n$ matched pairs
$\{\left((R'_{i1},S'_{i1}),(R'_{i2},S'_{i2})\right):i=1,\cdots,n\}$.
\item Output: RMSE for estimating the iROAS $\theta$
\item Procedure
\begin{enumerate}
\item Generate random geo assignment $\{A_{i},i=1,\cdots,n\}$, where $A_{i}$s
are i.i.d. with $P(A_{i}=1)=P(A_{i}=-1)=1/2$.
\item Generate experimental data as follows: if $A_{i}=-1$, then
\begin{align*}
X_{i}=r\cdot S'_{i1} & \text{ and }Y_{i}=(R'_{i1}+\theta\cdot r\cdot S'_{i1})-R'_{i2}
\end{align*}
otherwise $A_{i}=1$, then
\begin{align*}
X_{i}=r\cdot S'_{i2} & \text{ and }Y_{i}=(R'_{i2}+\theta\cdot r\cdot S'_{i2})-R'_{i1}
\end{align*}
where $r$ is chosen s.t. $\sum_{i=1}^{n}X_{i}=B$, i.e.
\begin{align*}
r & =\frac{B}{\sum_{i=1}^{n}\left(S'_{i1}I(A_{i}=-1)+S'_{i2}I(A_{i}=1)\right)}.
\end{align*}
\item Obtain the Trimmed Match point estimate $\hat{\theta}$ with experiment
data $\{(Y_{i},X_{i}),i=1,\cdots,n\}$ from b).
\item Replicate a)-c) for $K$ (say 1000) times to obtain $\{\hat{\theta}^{(k)},k=1,\cdots,K\}$;
The RMSE can then be approximated by
\begin{align*}
RMSE(\theta) & =\sqrt{\frac{1}{K}\sum_{k=1}^{K}(\hat{\theta}^{(k)}-\theta)^{2}}.
\end{align*}
\end{enumerate}
\end{itemize}
\end{algorithm}

As explained earlier, the magnitude of $\theta|S_{i1}-S_{i2}|$ is
usually much smaller than $|R_{i1}-R_{i2}|$, which results in $RMSE(\theta_{0})\approx RMSE(0)$,
or $RMSE$ for short. Then with the normal approximation, i.e. $\hat{\theta}\sim\mathcal{N}(\theta,RMSE^{2})$,
with Type-I error $\alpha$ and power $\beta$ for testing (\ref{eq:hypothesis-testing-framework}),
the classical statistical theory on hypothesis testing of normal mean
shift (\citet{bickel2015mathematical}) tells that
\begin{align}
\theta_{0} &= RMSE \times (q_{1-\alpha} + q_{\beta}),\label{eq:RMSE-theta0}
\end{align}
where $\theta_{0}$ is the minimum detectable iROAS defined in (\ref{eq:hypothesis-testing-framework}),
and $q_x$ is the $100\times x\%$ normal quantile.
Given $\theta_0$ and Type-I error $\alpha$, this tells that
for a desirable design, the RMSE must be no more
than $\theta_0 / (q_{1-\alpha} + q_{\beta})$ in order to have power no less than $\beta$.
\begin{rem}
\label{rem:permutation-test}Instead of using RMSE, one may
evaluate the power directly by
$$\frac{1}{K}\sum_{k=1}^{K}I(\hat{\theta}^{(k)}>q)$$
under $H_{1}$ at $\theta=\theta_{0}$, where $q$ is the $100\times (1-\alpha)\%$
quantile of $\hat{\theta}$ under
$H_{0}$ and can be approximated by the empirical quantile of $\hat{\theta}$
using simulations as described in Algorithm \ref{alg:RMSE-evaluation}. When the trim rate $\lambda$
in Trimmed Match is set to 0, then due to the fixed budget, $\hat{\theta}$
is proportional to the simulated response difference between treatment
and control. Therefore, the procedure is equivalent to the standard
permutation test.
\end{rem}

It is important to emphasize that if the pairing data during $\mathcal{T}_{p}$
is reused for the RMSE evaluation, RMSE may be under-estimated, similar
to the overfitting problem in machine learning. On the other hand,
with cross validation, the RMSE evaluation can give a reliable assessment
as long as the distribution of the underlying responses and spends during the test period is
similar to that during the evaluation period. When there
is a systematic difference between the evaluation period and the test
period, say due to the business growth, one may need to apply
some adjustment to RMSE.

Note that with a fixed budget, $r$ in Algorithm \ref{alg:RMSE-evaluation}
depends on the random assignment, and thus the response and spend data
for any geo pair will be affected by the treatment assignment to other
geo pairs. As a consequence, the simulation is incompatible with the
usual stable unit treatment value assumption for causal inference (c.f. \citet{imbens2015causal})
which assumes that the outcome on one unit should be unaffected by
the particular treatment assignment to other units. However, it may
be interesting to point out that the model (\ref{eq:model-iROAS})
still holds due to (\ref{eq:diff-Z-symmetry}), where geo assignment
affects ad spend $S_{ij}$ but not the ``uninfluenced response''
$Z_{ij}$. Therefore the power analysis based on the Trimmed Match
estimator is still valid.

\subsection{\label{subsec:Adjust-random-imbalance}Adjust random imbalance by
rerandomization}

After the construction of $n$ geo pairs, the two geos within each
pair are randomly assigned to treatment and control. While randomization
is considered the ``gold standard'', the treatment group and
the control group may not be perfectly comparable. As an extreme example,
with probability $2^{-n}$, all the $n$ geos with higher response
values within each pair will be assigned to the treatment group, which
would result in over-estimated iROAS. Such random imbalance should
be adjusted by rerandomizing the assignment until some balance checks
are met \citep{tukey1993tightening}, see \citet{morgan2012rerandomization}
for some general theory. Here are two of such checks: whether the
proportion of positive pairs (based on the sign of $R_{i1}-R_{i2}$)
is close to $1/2$, and whether the estimated iROAS with simulated
incremental ad spend is close to 0.

Asymptotically, rerandomization improves the estimation precision
(see \citet{li2020rerandomization}), in practice the sample
size is often small and it is still not clear how to perform balance
checks in an optimal way--too many balance checks may lead to very
few or even no options for geo assignment, which is not desirable.
A more rigorous investigation is desirable.

\subsection{Summary}

The entire design procedure can be summarized as follows:
\begin{itemize}
\item Split the pretest time period into two non-overlapping periods: a
time period for geo pairing ($\mathcal{T}_{p}$) and an evaluation
time period for power analysis ($\mathcal{T}_{e}$).
\item For each $n\leq N/2$, use data during $\mathcal{T}_{p}$ to generate
the optimal subset of $n$ geo pairs from $N$ geos as described in
Section \ref{subsec:optimal-pairing}. As a rule of thumb, we recommend
$n\geq10$ as too few pairs may make the inference unreliable.
\item For each $n$ and the optimal subset of $n$ pairs obtained in the
previous step, use data during $\mathcal{T}_{e}$ to perform power
analysis as described in Section \ref{subsec:Power-analysis}.
\item For each candidate set of geo pairs, randomize geo assignment within
each pair, and if needed adjust random imbalance through rerandomization
as discussed in Section \ref{subsec:Adjust-random-imbalance}.
\item Finally, choose the design which has small enough RMSE according to
the configuration for testing (\ref{eq:hypothesis-testing-framework}).
Here additional marketing constraints may apply, e.g. avoiding diminishing
returns if too much ad spend, limited ad inventory, etc.
\end{itemize}

\section{Simulation studies}\label{sec:simulations}

In this section, we use a simulated data set to illustrate how the design
works.

\subsection{Data simulation}

As a toy example, we first simulate geo sizes: for $i\in\{1,\cdots,N\}$,
\begin{align*}
g_{i} & =10^{5}\times F^{-1}\left(\frac{i}{N+1}\right)
\end{align*}
where $F^{-1}$ is the quantile function of log-normal with mean 1
and standard deviation 1 in the logarithmic scale, and then simulate
the daily response from a stationary time series below:
\begin{align*}
R_{i}(t) & =g_{i}\times\left(1+0.25\times w(t)\times(1+0.5\times\epsilon_{i}(t))\right)
\end{align*}
where $w(t)=\sin(2\pi t/7)$ simulates the typical day of the week
effect shared by all geos, and $\epsilon_{i}(t)=0.5\times\epsilon_{i}(t-1)+\mathcal{N}_{it}(0,1)$.
So the response is roughly proportional to the geo size, with some
fluctuation due to seasonality and some temporal AR(1) noise. The
spend proxy for the $i$th geo at time $t$ is simulated according
to
\begin{align}
S_{i}(t) & =0.01\times R_{i}(t)\times(1+0.5\times\mathcal{U}_{it}(-1,1)).\label{eq:spend-proxy}
\end{align}

In the above, $\mathcal{N}_{it}(0,1)$ and $\mathcal{U}_{it}(-1,1)$,
with $1\leq i\leq N$, denote mutually independent random numbers
generated from $\mathcal{N}(0,1)$ and $\mathcal{U}(-1,1)$ respectively.
All the parameters are specified explicitly for the ease of reproducibility.
\begin{center}
\begin{figure}
\begin{centering}
\includegraphics[scale=0.5]{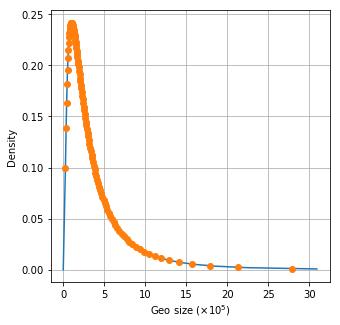}
\par\end{centering}
\caption{\label{fig:geo-size-lognormal}Geo sizes (circles) displayed on the lognormal density curve.}
\end{figure}
\par\end{center}

We simulate $N=100$ geos for $6$ weeks of pretest data. Figure \ref{fig:geo-size-lognormal}
shows the geo sizes $\{g_i, i=1,\cdots,N\}$ on the lognormal density curve.
Figure \ref{fig:Revenue-time-series}
provides a sample of response time series data for a few geos.
\begin{center}
\begin{figure}
\begin{centering}
\includegraphics[scale=0.5]{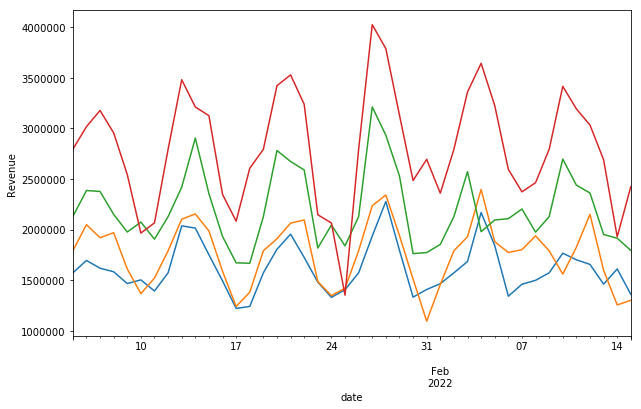}
\par\end{centering}
\caption{\label{fig:Revenue-time-series}Daily time series of the response
metric for a few sample geos.}
\end{figure}
\par\end{center}

\subsection{Result}

Consider designing a hold-back experiment with the budget $B=10^{6}$.
Suppose the test period is planned to be 2 weeks long and to start soon after the design.
The pretest data is split into two time periods: the most recent 4 weeks
of data is used for pairing ($\mathcal{T}_{p}$), and the first 2 weeks
for evaluation ($\mathcal{T}_{e}$). The purpose of using the latest
data for pairing is to best maintain the pairing quality from the
pairing period to the test period. By applying the design procedure
as described in Section \ref{sec:The-design-procedure}, for each
$n\in\{10,\cdots,50\}$, the corresponding optimal set of $n$ geo
pairs is generated according to the procedure in Section \ref{subsec:optimal-pairing},
and the RMSE value is obtained by Algorithm \ref{alg:RMSE-evaluation}.
Regarding the $L$ criteria, in the calculation of the distance (\ref{eq:geo-distance}),
$T$ spans one week.

The entire simulation is replicated for $M=5000$ times, that is, we use the above
data simulation procedure to generate 5000 pretest data sets and rerun the same
design process for each of them. Unless specified,
any reported result below is an average value across the 5000 replicates.

\subsubsection*{Effect of geo trimming}

With smaller $n$, more geos need to be excluded from the experiment, then we would expect
smaller RMSE as discussed in Section \ref{subsec:optimal-pairing}.
In Figure \ref{fig:RMSE-vs-n}, the x-axis is the number of geo pairs
$n$ and the y-axis on the left is the RMSE evaluated using data during
$\mathcal{T}_{e}$. As expected, the result shows that RMSE decreases
significantly as $n$ decreases. By reducing the number of pairs from
50 to 45, RMSE is reduced by nearly 50\%. The red curve (increasing) shows the
ratio of budget to the overall base response for the treatment group during $\mathcal{T}_{e}$,
which may be helpful for marketing decisions. For example, if $\theta_{0}=2.56$
and the type-I error and the power are set to 10\% and 90\% respectively,
it is desirable to have $RMSE\leq1$, for which Figure \ref{fig:RMSE-vs-n}
suggests $n<30$. On the other hand, from the marketing point of
view, if it is preferable that the budget-to-baseline ratio is no
more than, say 2\%, then the result suggest $n\geq 24$, and thus
the optimal design corresponds to $n=24$ as it minimizes
the RMSE and thus maximizes the power.
\begin{center}
\begin{figure}
\begin{centering}
\includegraphics[scale=0.5]{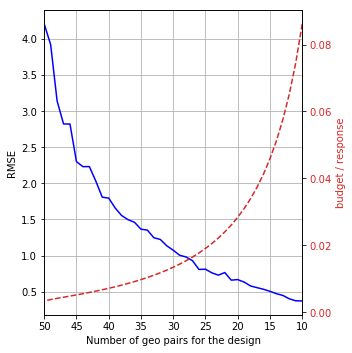}
\par\end{centering}
\caption{\label{fig:RMSE-vs-n}RMSE varies with the number of geo pairs, where
the red line (increasing) shows the corresponding ratio of budget
to baseline response.}
\end{figure}
\par\end{center}

\subsubsection*{Effect of cross validation}

\begin{figure}
\begin{centering}
\begin{minipage}[t]{0.5\columnwidth}%
\begin{center}
\includegraphics[scale=0.55]{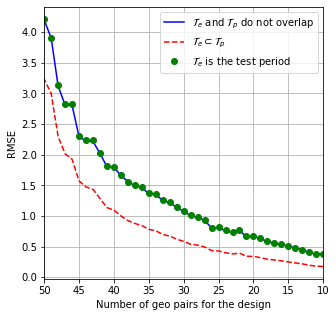}
\par\end{center}
\caption{\label{fig:Effect-of-sample-splitting}Effect of CV.}
\end{minipage}%
\begin{minipage}[t]{0.5\columnwidth}%
\begin{center}
\includegraphics[scale=0.55]{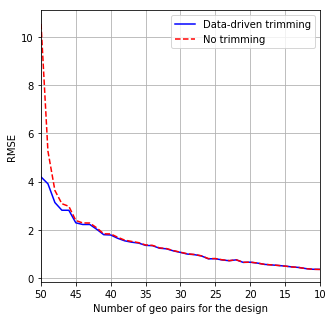}
\par\end{center}
\caption{\label{fig:RMSE-trim_rate_0}Effect of trimming.}
\end{minipage}
\par\end{centering}
\end{figure}

To see the effect of cross validation, the RMSE value is also evaluated using the most recent
2 weeks which falls into the pairing period $\mathcal{T}_{p}$. This
is shown as the dashed red curve in Figure \ref{fig:Effect-of-sample-splitting},
where the solid blue curve corresponds to $\mathcal{T}_{e}$ as reported
in Figure \ref{fig:RMSE-vs-n}. The result shows that the
RMSE value without cross validation is significantly lower than the corresponding
RMSE value with cross validation given the same number of geo pairs,
which confirms systematic overfitting.

Due to stationarity in the time series data generation, we expect the RMSE
evaluation to be unbiased. To verify that,
we have also extended the data generation from 6 weeks to 8 weeks and treated the 7th and 8th weeks
as the test period to reevaluate the RMSE for each of the 5000 replicates.
The average RMSE for each number of geo pairs is reported as the green circles on
Figure \ref{fig:Effect-of-sample-splitting}. Note that the green circles almost
overlap with the solid blue curve, which suggests that the RMSE
evaluation with cross validation is indeed unbiased.

\subsubsection*{Comparison with a design based on permutation test}

We also investigate the benefit of using Trimmed Match, over a design based on
the standard permutation test, which corresponds
to using Trimmed Match with a fixed $\lambda=0$ (see Remark \ref{rem:permutation-test}).
The result is summarized in Figure \ref{fig:RMSE-trim_rate_0}, which
shows interestingly that the performance is quite comparable except
for large $n$. In fact, when $n=50$, Trimmed Match reduces RMSE
by a factor larger than 2, which is not surprising though since
without trimming, log-normal is indeed much more heavy-tailed than
normal. As $n$ gets smaller, larger geos tend to be not matched as they
are harder to find comparable geos, and thus the remaining geos tend
to be more comparable to each other. Consequently, the response
differences tend to be less heavy-tailed, where the ideal trim rate
in Trimmed Match is close to 0. This not only validates the robustness
of the Trimmed Match estimator but provides some new insights specific
to the design process, which has not been studied in \citet{chen2019robust}.

\subsubsection*{Choice of the loss function}

\begin{figure}
\begin{centering}
\begin{minipage}[t]{0.5\columnwidth}%
\begin{center}
\includegraphics[scale=0.55]{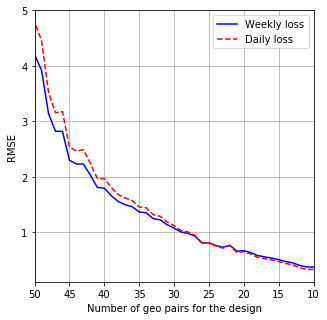}
\par\end{center}
\caption{\label{fig:daily-weekly-loss}Effect of $T$.}
\end{minipage}%
\begin{minipage}[t]{0.5\columnwidth}%
\begin{center}
\includegraphics[scale=0.55]{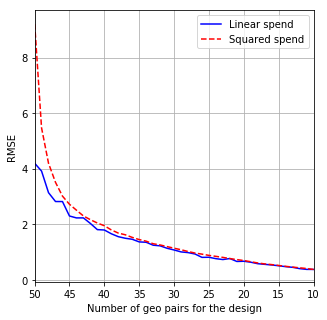}
\par\end{center}
\caption{\label{fig:sensitivity-spend}Effect of spend.}
\end{minipage}
\par\end{centering}
\end{figure}

We also report the performance comparison when the distance (\ref{eq:geo-distance}) uses
a day instead of a week for $T$, as shown in Figure \ref{fig:daily-weekly-loss}.
It is interesting to see that
the RMSE for $T$ being a week is smaller than that for $T$ being a day when $n$ is
large, while the inverse holds when $n$ is small. A heuristic explanation is that
for small $n$, the two geos within each pair are quite comparable--more
granular comparison (daily instead of weekly here) helps further differentiate geos which are
otherwise similar at a coarser granularity.

\subsubsection*{Sensitivity to the spend proxy}

The power calculation of the design is determined by the baseline
response and the spend proxy used in the RMSE evaluation. Here we
illustrate this by looking at different choices of spend proxy.
Note that in the above simulation, the spend proxy (\ref{eq:spend-proxy})
is proportional to a linear scale of the response, which is often
expected. As a comparison, we replicate the above simulation studies,
except that the spend proxy in (\ref{eq:spend-proxy}) is now taken
to be proportional to the squared response (which is very extreme)
as follows:
\begin{align}
S_{i}(t) & \propto\left(R_{i}(t)\right)^{2}\times(1+0.5\times\mathcal{U}_{it}(-1,1)).\label{eq:squared-spend-proxy}
\end{align}
The RMSE curve as a function of the number of geo pairs for the design
is reported by the dotted red line in Figure \ref{fig:sensitivity-spend},
whereas the solid blue line is based on the spend
proxy (\ref{eq:spend-proxy}) as reported earlier. The result shows
that when $n$ is close to 50, the squared spend leads to generally higher RMSEs than
the linear spend, which can be explained briefly as follows: for
the squared spend, too much budget would be spent on the largest geo
pair, where the pairing quality is the worst
(e.g. the size difference of the largest two geos which makes a pair
is more than twice the median geo size, see Figure \ref{fig:geo-size-lognormal}),
while for the linear spend, the budget is more evenly distributed to geo pairs
with better pairing quality, which helps reduce the RMSE.
On the other hand, the result also shows that for smaller $n$ (say less than 40),
the RMSEs between the linear spend and the squared spend are quite comparable,
which suggests that the method is quite robust against potential misspecification
of the spend proxy.

\section{A case study}\label{sec:real-case-study}

\begin{figure}
\begin{centering}
\includegraphics[scale=0.5]{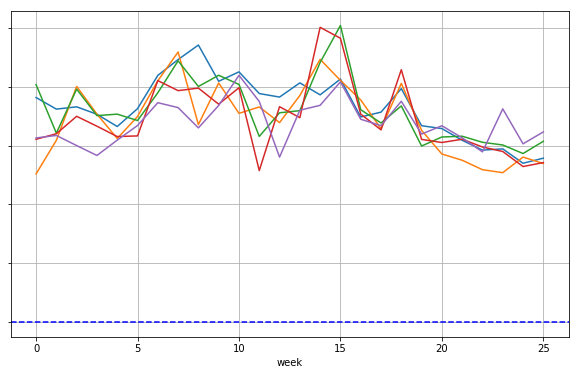}
\par\end{centering}
\caption{\label{fig:Revenue-time-series-real}Time series of the weekly revenue
data for a few geos, where the horizontal dashed line has value 0
while the revenues are jittered and the detailed dates and revenue scales are removed in order to
anonymize the experiment.}
\end{figure}

In this section we report a real case study, where the experiment
was to test a new type of ad. Nielsen's 210 DMAs were used as the
geos, except that a few DMAs were excluded for some other marketing
purpose. To design the experiment, more than 1 year of historical
revenue (including both offline and online sales) data $R_{g}(t)$
were collected for each geo on a daily basis. To give an idea of
the real data, Figure \ref{fig:Revenue-time-series-real} shows the
weekly time series of $R_{g}(t)\cdot(1+c\cdot U_{g}(t))$ for a few
sample geos with both the dates and detailed scales removed, where
$\{U_{g}(t):g,t\}$ i.i.d uniformly distributed on $(-1,1)$ and a
small constant $c\in(0,1)$ are used to further anonymize the data.
The figure shows that unlike the simulated data, the real revenue
here is quite non-stationary.

\begin{figure}
\begin{centering}
\includegraphics[scale=0.5]{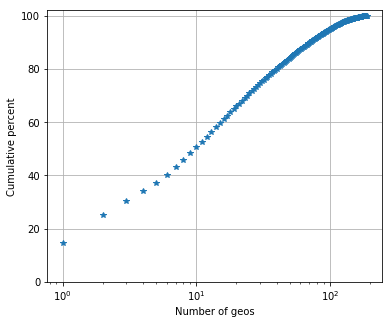}
\par\end{centering}
\caption{\label{fig:revenue-power-law-real}Geo-level revenues approximately follow a Pareto distribution.}
\end{figure}

To illustrate the geo heterogeneity, we calculate the aggregated revenue
for each geo, sort the geos decreasingly based on the aggregated revenues,
and then look at the cumulative revenues of geos,
divided by total revenue, which is shown in Figure \ref{fig:revenue-power-law-real},
where the x-axis shows the corresponding number of geos in the logarithmic scale.
The figure shows that the geo sizes measured by revenues approximately
follow a heavy-tailed Pareto distribution\footnote{\url{https://en.wikipedia.org/wiki/Pareto_distribution}},
with about 80\% of the revenue from 20\% of the geos. Some historical
ad spend data were also collected to inform the split of the client
pre-specified budget.

\begin{figure}
\centering{}%
\begin{minipage}[t]{0.5\columnwidth}%
\begin{center}
\includegraphics[scale=0.55]{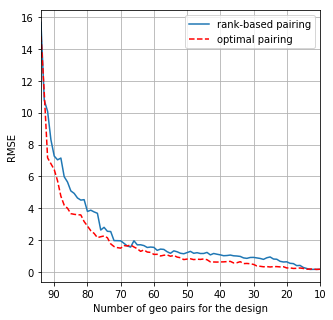}
\par\end{center}
\caption{\label{fig:Rank-vs-optimal-rmse}Rank-based pairing vs optimal pairing}
\end{minipage}%
\begin{minipage}[t]{0.5\columnwidth}%
\begin{center}
\includegraphics[scale=0.55]{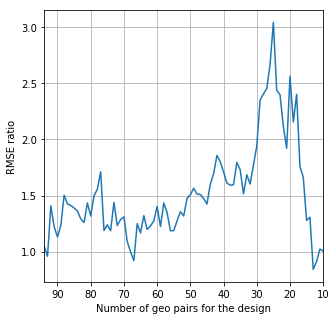}
\par\end{center}
\caption{\label{fig:RMSE-ratio}RMSE ratio between rank-based pairing and optimal
pairing}
\end{minipage}
\end{figure}

The experiment was planned for 6 weeks of testing, plus 2 weeks of cooldown to capture
any potential lag effects. The evaluation period was taken to be 8 weeks so as
to have the same duration as the experiment,
and by looking at the seasonality and year-to-year patterns, the 8
weeks were chosen to be the same weeks of the year as the planned
experimentation period, but in the previous year. The data after that is used
for geo pairing. We investigated two pairing algorithms: optimal pairing
as described in Section \ref{subsec:optimal-pairing} where $T$ is
a week, and the rank method which ranks and pairs the geos based on
their revenue volumes during the pairing period as described in Remark
\ref{rem:rank-based-pairing}. The RMSE curve as a function of the
number of geo pairs is shown in Figure \ref{fig:Rank-vs-optimal-rmse}
for the rank-based pairing (solid blue) and optimal pairing (dashed
red) separately. The result tells that, in general, optimal pairing
performs significantly better than the simpler rank-based method.
Figure \ref{fig:RMSE-ratio} further plots the ratio of the RMSEs
between the two methods for each number of geo pairs and shows that
the reduction of RMSE by using optimal pairing compared to the rank-based
pairing can be as large as a factor of 3.

The actual design which was employed for the experiment followed the
procedure as described in Section \ref{sec:The-design-procedure},
except that the pairing algorithm was based on the rank method as
optimal pairing was not implemented at that time yet. It used around
60 pairs by applying the budget and other marketing constraints. After
the experiment was completed, the point estimate and confidence interval
were obtained by the Trimmed Match estimator (see \citet{chen2019robust}
for the detailed description). It may be interesting to point out
that the half width of the $80\%$ two-sided confidence interval is
quite comparable to $1.28\times RMSE$, where $RMSE$ was reported
from the design phase, despite the irregular temporal dynamics of
the revenues.

\section{Conclusion}

We have proposed a robust and cost-effective method for designing
randomized paired geo experiments for measuring the incremental return
on ad spend. The method integrates optimal pairing, Trimmed Match for estimating
iROAS,
and cross validation in a novel and systematic manner in order to address
both geo heterogeneity and temporal dynamics. The method includes
the standard permutation test as a special example when the trim rate in the Trimmed
Match estimator is set to 0. By trimming a few geos and constructing an optimal subset
of pairs in a data-driven manner, the new method can often reduce the
cost by a large factor in real studies. Various open problems exist, such as
how to improve the $L$ criterion by taking into account the uncertainty in the data-driven
choice of trim rate,
how to address random imbalance in a more rigorous manner,
and how to adjust the RMSE due to the potential difference between the evaluation period and the test period.

\section*{Acknowledgments}

We would like to thank Jim Koehler, Tim Au, Jouni Kerman, Kevin Benac, Fan Zhang,
Christoph Best, Shu Li, Ricardo Marino, Jim Dravillas, Thomas Kondrat, Andree Lischewski,
Yan Sun, Xiaoyue Zhao for insightful technical discussions, Tony Fagan, Penny Chu, Isabel Marcin, Kate
O'Donovan, Manojav Patil, Arthur Anglade, Rachel Fan, Irene Nocon,
Karina Przyjemski, and numerous colleages in AMT, Eng, PM and the sales teams for
the support and case studies.

\bibliographystyle{plainnat}
\bibliography{trimmed_match_design}

\end{document}